\documentclass[epj]{svjour}
\usepackage{latexsym}
\usepackage{amsmath,amssymb,amsfonts,wasysym}
\usepackage{graphics}
\usepackage{xcolor}
\usepackage{float}
\usepackage{hyperref}
\usepackage{longtable}
\usepackage{verbatim}
\usepackage{cite}

\numberwithin{equation}{section}

\def\({\left(}
\def\){\right)}

\def\be{\begin{equation}}
\def\ee{\end{equation}}

\input{tcilatex}
\begin{document}

\title{$S_3$ discrete group as a source of the quark mass and mixing pattern in $331$ models.}
\author{A. E. C\'arcamo Hern\'andez  \thanks{antonio.carcamo@usm.cl}, R. Martinez \thanks{remartinezm@unal.edu.co} and Jorge Nisperuza \ {$^{c}$} 
}
\institute{$^{a}$Universidad T\'ecnica Federico Santa Mar\'{\i}a and Centro Cient\'{\i}%
fico-Tecnol\'ogico de Valpara\'{\i}so, Casilla 110-V, Valpara\'{\i}so,
Chile, $^{{b,c}}$Departamento de F\'{\i}sica, Universidad Nacional de
Colombia,Ciudad Universitaria, Bogot\'{a} D.C., Colombia.}
\date{Received: date / Revised version: date}

\abstract{We propose a model based on the $SU(3)_{C}\otimes SU(3)_{L}\otimes U(1)_{X}$
gauge symmetry with an extra $S_{3}\otimes Z_{2}\otimes Z_{4}\otimes Z_{12}$
discrete group, which successfully accounts for the SM quark mass and mixing pattern.
The observed hierarchy of the SM quark masses and quark mixing matrix elements
arises from the $Z_{4}$ and $Z_{12}$ symmetries, which are broken at very
high scale by the $SU(3)_{L}$ scalar singlets ($\sigma$,$\zeta$) and $\tau $, charged
under these symmetries, respectively. The Cabbibo mixing arises from the
down type quark sector whereas the up quark sector generates the
remaining quark mixing angles. The obtained magnitudes of the CKM matrix elements, the CP violating phase and the Jarlskog invariant are in agreement with the
experimental data.}

\PACS{
   {PACS-key}{discribing text of that key}   \and
    {PACS-key}{discribing text of that key}
    } 
%
%

\authorrunning{A. C\'arcamo, R. Martinez, J. Nisperuza} 
\titlerunning{Quark
masses and mixings in 331S3}

\maketitle

\section{Introduction}

\label{intro}The discovery of a scalar field with a mass of $125$ GeV by LHC
experiments \cite{atlashiggs,cmshiggs,newtevatron,CMS-PAS-HIG-12-020}
confirms that the Standard Model (SM) is the right theory of electroweak
interactions and may provide an explanation for the origin of mass of
fundamental particles and for the spontaneous symmetry breaking. Despite the
success of the LHC experiments, there are many aspects not yet explained
such as the fermion mass hierarchy. This discovery of the Higgs scalar field
opens the possibility to formulate theories beyond the SM that include
additional scalar fields that can be useful to explain the existence of Dark
Matter \cite{BSMtheorieswithDM}.

One of the outstanding unresolved issues in Particle Physics is the origin
of the masses of fundamental fermions. The current theory of strong and
electroweak interactions, the Standard Model (SM), has proven to be
remarkably successful in passing all experimental tests. Despite its great
success, the Standard Model (SM) based on the $SU(3)_{C}\otimes
SU(2)_{L}\otimes U(1)_{Y}$ gauge symmetry is unlikely to be a truly
fundamental theory due to unexplained features \cite{SM,PDG}. Most of them
are linked to the existence of three families of fermions as well as the
fermion mass and mixing hierarchy; problems presented in its quark and
lepton sectors. Neutrino oscillation experiments provide a clear indication
that neutrinos are massive particles, but these experiments do not explain
neither the neutrino mass squared splittings nor the Dirac or Majorana
identity of neutrinos
. While in the quark sector the mixing angles are small, in the lepton
sector two of the mixing angles are large, and one mixing angle is small.
This suggests different mechanisms for the generation of mass in the quark
and lepton sectors. Experiments with solar, atmospheric and reactor
neutrinos provide evidence of neutrino oscillations from the measured non
vanishing neutrino mass squared splittings.

One clear and outstanding feature in the pattern of quark masses is that
they increase from one generation to the next spreading over a range of five
orders of magnitude \cite{PDG,Fritzsch,Xing2011}. From the phenomenological
point of view, it is possible to describe some features of the mass
hierarchy by assuming zero-texture Yukawa matrices \cite{textures,GUT,Extradim,String}.
Recently, discrete groups have been considered to explain the observed
pattern of fermion masses and mixing \cite{Altarelli:2010gt,Ishimori:2010au,King:2013eh,discrete-lepton,discrete-quark,s3pheno,Delta27,Tprime,Hernandez:2015tna,Hernandez:2015cra}. Other models with
horizontal symmetries have been proposed in the literature \cite{horizontal}.

On the other hand, the origin of the structure of fermions can be addressed
in family dependent models. Alternatively, an explanation to this issue can
also be provided by the models based on the gauge symmetry $SU(3)_{c}\otimes
SU(3)_{L}\otimes U(1)_{X}$, also called 3-3-1 models, which introduce a
family non-universal $U(1)_{X}$ symmetry \cite{331-pisano,331-frampton,331-long,M-O}. Models based on the gauge symmetry $%
SU(3)_{C}\times SU(3)_{L}\times U(1)_{X}$ are very interesting since they
predict the existence of three families from the quiral anomaly cancellation 
\cite{anomalias}. In these models, two families of quarks have the same
quantum numbers, which are associated to the two families of light quarks to
correctly predict the Cabbibo mixing angle. The third family has different $%
U(1)_{X}$ values and thus it is associated to the heavy quarks. Thus, the
fact that the third family is treated under a different representation, can
explain the large mass difference between the heaviest quark family and the
two lighter ones \cite{third-family}. These models include a Peccei-Quinn
symmetry that sheds light into the strong CP problem \cite{PC}. The 331
models with sterile neutrinos have weakly interacting massive fermionic dark
matter candidates \cite{DM331}.

In this paper we propose a version of the $SU(3)_{C}\times SU(3)_{L}\times
U(1)_{X}$ model with an additional discrete symmetry group 
\mbox{$S_{3}\otimes
Z_{2}\otimes Z_{4}\otimes Z_{12}$} and an extended scalar sector needed in
order to reproduce the specific patterns of mass matrices in the quark
sector that successfully account for the quark mass and mixing hierarchy.
The particular role of each additional scalar field and the corresponding
particle assignments under the symmetry group of the model are explained in
details in Sec. \ref{model}. Our model successfully describes the prevailing
pattern of the SM quark masses and mixing.

This paper is organized as follows. In Sec. \ref{model} we outline the
proposed model. In Sec. \ref{quarkmassesandmixing} we present our results in
terms of quark masses and mixing, which is followed by a numerical analysis.
In Sec. \ref{scalarpotential}, we discuss the scalar mass spectrum resulting
from the low energy scalar potential. Finally in Sec. \ref{conclusions}, we
state our conclusions. In the appendixes we present several technical
details. Appendix \ref{ap0} gives a brief description of the $S_{3}$ group.
Appendix \ref{ap2} presents a discussion of the stability conditions of the
low energy scalar potential.

\section{The Model}

\label{model} We consider an extension of the minimal $SU(3)_{C}\otimes
SU\left( 3\right) _{L}\otimes U\left( 1\right) _{X}$ (331) model with the
full symmetry $\mathcal{G}$ experiencing a three-step spontaneous breaking: 
\begin{eqnarray}
&&\mathcal{G}=SU(3)_{C}\otimes SU\left( 3\right) _{L}\otimes U\left(
1\right) _{X}\otimes S_{3}\otimes Z_{2}\otimes Z_{4}\otimes Z_{12}  \notag \\
&&\hspace{35mm}\Downarrow \Lambda _{int}  \notag \\[0.12in]
&&\hspace{15mm}SU(3)_{C}\otimes SU\left( 3\right) _{L}\otimes U\left(
1\right) _{X}  \notag \\[0.12in]
&&\hspace{35mm}\Downarrow v_{\chi }  \notag \\[0.12in]
&&\hspace{15mm}SU(3)_{C}\otimes SU\left( 2\right) _{L}\otimes U\left(
1\right) _{Y}  \notag \\[0.12in]
&&\hspace{35mm}\Downarrow v_{\eta },v_{\rho }  \notag \\[0.12in]
&&\hspace{15mm}SU(3)_{C}\otimes U\left( 1\right) _{Q}  \label{Group}
\end{eqnarray}%
where the different symmetry breaking scales satisfy the following hierarchy 
$\Lambda _{int}\gg v_{\chi }\gg v_{\eta },v_{\rho }.$ 

In our model 331 model, the electric charge is defined in terms of the $%
SU(3) $ generators and the identity by: 
\begin{equation}
Q=T_{3}-\frac{1}{\sqrt{3}}T_{8}+XI,
\end{equation}%
with $I=Diag(1,1,1)$, $T_{3}=\frac{1}{2}Diag(1,-1,0)$ and $T_{8}=(\frac{1}{2%
\sqrt{3}})Diag(1,1,-2)$.

The anomaly cancellation of $SU(3)_{L}$ requires that the two families of
quarks be accommodated in $3^{\ast }$ irreducible representations (irreps).
From the quark colors, it follows that the number of $3^{\ast }$ irreducible
representations is six. The other family of quarks is accommodated with its
three colors, into a $3$ irreducible representation. When including the three families
of leptons, we have six $3$ irreps. Consequently, the $SU(3)_{L}$
representations are vector like and anomaly free. In order to have anomaly
free $U(1)_{X}$ representations, one needs to assign quantum numbers to the
fermion families in such a way that the combination of the $U(1)_{X}$
representations with other gauge sectors be anomaly free. Therefore, from
the requirement of anomaly cancellation we get the following $%
(SU(3)_{C},SU(3)_{L},U(1)_{X})$ left handed fermionic representations:

\begin{align}
Q_{L}^{1,2}& =%
\begin{pmatrix}
D^{1,2} \\ 
-U^{1,2} \\ 
J^{1,2} \\ 
\end{pmatrix}%
_{L}:(3,3^{\ast },0),\hspace{0.5cm}  \notag \\
Q_{L}^{3}& =%
\begin{pmatrix}
U^{3} \\ 
D^{3} \\ 
T \\ 
\end{pmatrix}%
_{L}:(3,3,1/3),  \notag \\
L_{L}^{1,2,3}& =%
\begin{pmatrix}
\nu ^{1,2,3} \\ 
e^{1,2,3} \\ 
(\nu ^{1,2,3})^{c} \\ 
\end{pmatrix}%
_{L}:(1,3,-1/3),  \label{fermion_spectrum}
\end{align}%
%
%
%
%
Let's note that the right-handed sector transforms as singlets under $SU(3)_{L}$. The right
handed up and down type SM quarks transform under $%
(SU(3)_{C},SU(3)_{L},U(1)_{X})$ as\ $U_{R}^{1,2,3}:(3^{\ast },1,2/3)$ and $%
D_{R}^{1,2,3}:(3^{\ast },1,-1/3)$, respectively. In addition, we see that
the model has the following heavy fermions: a single flavor quark $T$ with
electric charge $2/3$, two flavor quarks $J^{1,2}$ with charge $-1/3$. The
right handed sector of the exotic quarks transforms as $T_{R}:(3^{\ast
},1,2/3)$ and $J_{R}^{1,2}:(3^{\ast },1,-1/3)$. In the concerning to the
lepton sector, we have three right handed charged leptons $e_{R}^{1,2,3}:(1,1,-1)$ and three
right-handed Majorana leptons $N_{R}^{1,2,3}:(1,1,0)$ (recently, a
discussion about neutrino masses via double and inverse see-saw mechanism
was perform in ref. \cite{catano}).

The scalar sector of the 331 model includes three $3$'s irreps of $SU(3)_{L}$%
, where one triplet $\chi $ acquires a vacuum expectation value (VEV) at
high energy scale, $v_{\chi }$, responsible for the breaking of the $%
SU(3)_{L}\times U(1)_{X}$ symmetry down to the $SU(2)_{L}\times U(1)_{Y}$
electroweak group of the SM; and two light triplet fields $\eta $ and $\rho $
get VEVs $v_{\eta }$ and $v_{\rho }$, respectively, at the electroweak scale
and give mass to the fermion and gauge sector. In addition to the
aforementioned scalar spectrum, we introduce six $SU\left( 3\right) _{L}$
scalar singlets, namely, $\xi _{1}$, $\xi _{2}$, $\zeta _{1}$, $\zeta _{2}$, 
$\sigma $ and $\tau $. Their role and importance will be explained later in
this section.

The $[SU(3)_{L},U(1)_{X}]$ group structure of the scalar fields of our model
is: 
\begin{align}
\chi & =%
\begin{pmatrix}
\chi _{1}^{0} \\ 
\chi _{2}^{-} \\ 
\frac{1}{\sqrt{2}}(\upsilon _{\chi }+\xi _{\chi }\pm i\zeta _{\chi }) \\ 
\end{pmatrix}%
:(3,-1/3),\hspace{1cm}\xi _{1}:(1,0),\hspace{1cm}  \notag \\
\rho & =%
\begin{pmatrix}
\rho _{1}^{+} \\ 
\frac{1}{\sqrt{2}}(\upsilon _{\rho }+\xi _{\rho }\pm i\zeta _{\rho }) \\ 
\rho _{3}^{+} \\ 
\end{pmatrix}%
:(3,2/3),\hspace{1cm}\xi _{2}:(1,0),  \notag \\
\eta & =%
\begin{pmatrix}
\frac{1}{\sqrt{2}}(\upsilon _{\eta }+\xi _{\eta }\pm i\zeta _{\eta }) \\ 
\eta _{2}^{-} \\ 
\eta _{3}^{0}%
\end{pmatrix}%
:(3,-1/3),\hspace{1cm}\sigma :(1,0),  \notag \\
\tau & :(1,0),\hspace{1cm}\zeta _{1}:(1,0),\hspace{1cm}\zeta _{2}:(1,0).
\label{331-scalar}
\end{align}

We group the scalar fields into doublet and singlet representions of $S_{3}$. The $S_{3}\otimes Z_{2}\otimes Z_{4}\otimes Z_{12}$ assignments of the
scalar fields are:

\begin{eqnarray}
\Phi  &=&\left( \eta ,\chi \right) \sim \left( \mathbf{2,}1,1,1\right) ,%
\hspace{1cm}\rho \sim \left( \mathbf{1}^{\prime }\mathbf{,}1,1,1\right) ,%
\hspace{1cm}  \notag \\
\sigma  &\sim &\left( \mathbf{1,-}1,i,1\right) ,\hspace{1cm}\tau \sim \left( 
\mathbf{1,}1,1,\omega ^{-\frac{1}{4}}\right) ,  \notag \\
\xi  &=&\left( \xi _{1},\xi _{2}\right) \sim \left( \mathbf{2,-}1,1,1\right)
, \\
\zeta  &=&\left( \zeta _{1},\zeta _{2}\right) \sim \left( \mathbf{2,-}%
1,i,1\right) .
\end{eqnarray}%
where $\omega =e^{2\pi i/3}$.

Regarding the quark sector, we assign the quark fields in trivial and non
trivial singlet representions of $S_{3}$. We assumed that all left handed
quarks and right handed quarks are assigned to $S_{3}$ trivial singlets
excepting, $U_{R}^{1}$, $U_{R}^{2}$, $U_{R}^{3}$, $T_{R}$, $D_{R}^{3}$, $%
J_{R}^{1}$ and $J_{R}^{2}$, which are assumed to be non trivial singlets.
The quark assignments under $S_{3}\otimes Z_{2}\otimes Z_{4}\otimes Z_{12}$
are:

\begin{eqnarray}
Q_{L}^{1} &\sim &\left( \mathbf{1,}1,1,\omega ^{-\frac{1}{2}}\right) ,%
\hspace{1cm}Q_{L}^{2}\sim \left( \mathbf{1,}1,1,\omega ^{-\frac{1}{4}%
}\right) ,\hspace{1cm}  \notag \\
Q_{L}^{3} &\sim &\left( \mathbf{1,}1,1,1\right) ,\hspace{1cm}\hspace{1cm}%
U_{R}^{1}\sim \left( \mathbf{1}^{\prime }\mathbf{,}1,-1,\omega \right) , 
\notag \\
U_{R}^{2} &\sim &\left( \mathbf{1}^{\prime }\mathbf{,}1,-1,\omega ^{\frac{1}{%
4}}\right) ,\hspace{1cm}U_{R}^{3}\sim \left( \mathbf{1}^{\prime }\mathbf{,-}%
1,-i,1\right) , \\
D_{R}^{1} &\sim &\left( \mathbf{1,-}1,1,\omega \right) ,\hspace{1cm}%
D_{R}^{2}\sim \left( \mathbf{1,-}1,1,i\right) ,\hspace{1cm}  \notag \\
D_{R}^{3} &\sim &\left( \mathbf{1}^{\prime }\mathbf{,}1,1,i\right) ,\hspace{%
1cm}T_{R}\sim \left( \mathbf{1}^{\prime }\mathbf{,-}1,1,1\right) ,  \notag \\
J_{R}^{1} &\sim &\left( \mathbf{1}^{\prime }\mathbf{,-}1,1,\omega ^{-\frac{1%
}{2}}\right) ,\hspace{1cm}J_{R}^{2}\sim \left( \mathbf{1}^{\prime }\mathbf{,-%
}1,1,\omega ^{-\frac{1}{4}}\right) .  \notag
\end{eqnarray}

With the above particle content, the following relevant Yukawa terms for the
quark sector arise: 
\begin{eqnarray}
-\mathcal{L}_{Y} &=&y_{11}^{\left( U\right) }\overline{Q}_{L}^{1}\rho ^{\ast
}U_{R}^{1}\frac{\sigma ^{2}\tau ^{6}}{\Lambda ^{8}}+y_{22}^{\left( U\right) }%
\overline{Q}_{L}^{2}\rho ^{\ast }U_{R}^{2}\frac{\sigma ^{2}\tau ^{2}}{%
\Lambda ^{4}}  \notag \\
&&+y_{13}^{\left( U\right) }\overline{Q}_{L}^{1}\rho ^{\ast }U_{R}^{3}\frac{%
\sigma \tau ^{2}}{\Lambda ^{3}}+y_{23}^{\left( U\right) }\overline{Q}%
_{L}^{2}\rho ^{\ast }U_{R}^{3}\frac{\sigma \tau }{\Lambda ^{2}}  \notag \\
&&+y_{33}^{\left( U\right) }\overline{Q}_{L}^{3}\Phi U_{R}^{3}\frac{\zeta }{%
\Lambda }+y^{\left( T\right) }\overline{Q}_{L}^{3}\Phi T_{R}\frac{\xi }{%
\Lambda }  \notag \\
&&+y_{11}^{\left( D\right) }\overline{Q}_{L}^{1}\Phi ^{\ast }D_{R}^{1}\frac{%
\xi \tau ^{6}}{\Lambda ^{7}}+y_{12}^{\left( D\right) }\overline{Q}%
_{L}^{1}\Phi ^{\ast }D_{R}^{2}\frac{\xi \tau ^{5}}{\Lambda ^{6}}  \notag \\
&&+y_{22}^{\left( D\right) }\overline{Q}_{L}^{2}\Phi ^{\ast }D_{R}^{2}\frac{%
\xi \tau ^{4}}{\Lambda ^{5}}+y_{21}^{\left( D\right) }\overline{Q}%
_{L}^{2}\Phi ^{\ast }D_{R}^{1}\frac{\xi \tau ^{5}}{\Lambda ^{6}}  \notag \\
&&+y_{33}^{\left( D\right) }\overline{Q}_{L}^{3}\rho D_{R}^{3}\frac{\tau ^{3}%
}{\Lambda ^{3}}+y_{1}^{\left( J\right) }\overline{Q}_{L}^{1}\Phi ^{\ast
}J_{R}^{1}\frac{\xi }{\Lambda }  \notag \\
&&+y_{2}^{\left( J\right) }\overline{Q}_{L}^{2}\Phi ^{\ast }J_{R}^{2}\frac{%
\xi }{\Lambda }  \label{Yukawaterms}
\end{eqnarray}
where the dimensionless couplings $y_{ij}^{\left( U,D\right) }$ ($i,j=1,2,3$), $y^{(T)}$, $y^{(J)}_{1,2}$ are $\mathcal{O}(1)$ parameters.

To explain the fermion mass hierarchy it is necessary to assume an ansatz for
the Yukawa matrices. A candidate for generating specific Yukawa textures is
the $S_{3}\otimes Z_{2}\otimes Z_{4}\otimes Z_{12}$ discrete group that can
explain the prevailing pattern of fermion masses and mixing. The $S_{3}$ discrete symmetry is the smallest non-Abelian discrete symmetry group having
three irreducible representations (irreps), explicitly two singlets and one
doublet irreps. The $Z_{2}$ symmetry determines the allowed Yukawa terms for
the quark sector%
, thus resulting in a reduction of model parameters and allowing one to decouple
the bottom quark from the light down and strange quarks. The $Z_{4}$ and $Z_{12}$ symmetries shape the hierarchical structure of the quark mass matrices that yields a realistic pattern of quark masses and mixing. It is noteworthy that the properties of the $Z_{N}$ groups imply that the $Z_{4}$ and $Z_{12}$ symmetries are the lowest cyclic symmetries that allow one to buid Yukawa terms of dimensions six and ten, respectively. Consequently, the $Z_{4}\otimes Z_{12}$ symmetry is the
lowest cyclic symmetry from which a 12 dimensional Yukawa term can be
built, crucial to get the required $\lambda ^{8}$ supression in the 11 entry
of the up-type quark mass matrix, where $\lambda =0.225$ is one of the
Wolfenstein parameters. Furthermore, thanks to the $Z_{4}\otimes Z_{12}$ symmetry, the lowest down-type quark Yukawa term contributing to the 11
entry of the down-type quark mass matrix has dimension 11. Thus, the $%
Z_{4}$ and $Z_{12}$ symmetries are crucial to explain the smallness of the
up and down quark masses. 

We assume the following VEV pattern for the $SU\left( 3\right) _{L}$ singlet
scalar fields:
\begin{eqnarray}
\left\langle \xi \right\rangle  &=&v_{\xi }\left( 1,0\right) ,\hspace{1cm}%
\left\langle \tau \right\rangle =v_{\tau }  \notag \\
\left\langle \zeta \right\rangle  &=&v_{\zeta }\left( 0,1\right) ,\hspace{1cm%
}\left\langle \sigma \right\rangle =v_{\sigma }\label{VEV}.  
\end{eqnarray}%
i.e. the VEVs of $\xi$ and $\zeta$ are aligned as $(1,0)$ and $(0,1)$ in the $S_{3}$ directions, respectively. Besides that, the $SU\left( 3\right) _{L}$ scalar singlets, $\xi _{1}$, $\zeta_2$, $\sigma $ and $\tau $, are assumed to acquire VEVs at a scale $\Lambda _{int}$ much larger than $\upsilon _{\chi }$ in order to break the $\mathcal{G}=SU(3)_{C}\otimes SU\left( 3\right) _{L}\otimes U\left( 1\right) _{X}\otimes S_{3}\otimes Z_{2}\otimes Z_{4}\otimes Z_{12}$ symmetry group down to $%
SU(3)_{C}\otimes SU\left( 3\right) _{L}\otimes U\left( 1\right) _{X}$. Let us note that the $S_{3}$ doublet $SU\left( 3\right) _{L}$ singlet scalars $\xi$ and $\zeta$ are the only scalar fields odd under the $Z_{2}$ symmetry. Furthermore the only scalar fields charged under the $Z_{4}$ and $Z_{12}$ symmetries are the 
$SU\left( 3\right) _{L}$ singlet scalars $\sigma$, $\zeta$ and $\tau $,
respectively. Thus, the breaking of the $Z_{2}$, $Z_{4}$ and $Z_{12}$
symmetries is caused by the scalar fields $\xi $, ($\sigma$, $\zeta$) and $\tau $, respectively, acquiring VEVs at a very high scale. It is worth mentioning that we have chosen a VEV patterns for the $S_{3}$ doublets $SU\left( 3\right) _{L}$
singlet scalar $\xi $ and $\zeta$, in the $\left( 1,0\right)$ and $\left( 0,1\right)$ $S_{3}$ directions, respectively, as indicated by Eq. (\ref{VEV}), in order to decouple the heavy exotic quarks from the SM quarks. Due to the aforementioned choice of the VEV pattern of $\xi $, only the $SU(3)_{L}$ scalar triplet $\chi $ participates in the Yukawa interactions giving masses to the exotic $T$, $J^{1}$ and $J^{2}$ quarks. Furthermore, the masses of the SM quarks will arise from the Yukawa terms involving the $SU(3)_{L}$ scalar triplets $\eta $ and $\rho $.

Considering that the quark mass and mixing pattern arises from the $Z_{4}$
and $Z_{12}$ symmetries, and in order to relate the quark masses with the
quark mixing parameters, we set the VEVs of the $SU\left( 3\right) _{L}$
singlet scalar fields excepting $v_{\zeta }$ as follows:

\begin{equation}
v_{\xi }=v_{\tau }=v_{\sigma }=\Lambda _{int}=\lambda \Lambda ,
\label{VEVsinglets}
\end{equation}
where $\lambda =0.225$ is one of the parameters of the Wolfenstein
parametrization and $\Lambda $ is the cutoff of our model.

To reproduce the right value of the top quark mass while keeping $%
y_{33}^{\left( U\right) }\sim \mathcal{O}(1)\lesssim \sqrt{4\pi }$ as
required by perturbativity, we set $v_{\zeta }$ in the following range:

\begin{equation}
\frac{\sqrt{2}m_{t}}{\sqrt{4\pi }v_{\eta }}\Lambda <v_{\zeta }<\Lambda .
\label{vzeta}
\end{equation}

\section{Quark masses and mixing.}

\label{quarkmassesandmixing} Using Eq. (\ref{Yukawaterms}) and considering
that the VEV pattern of the $SU\left( 3\right) _{L}$ singlet scalar fields
satisfies Eq. (\ref{VEV}) with the nonvanishing VEVs set to be equal to $%
\lambda \Lambda $ (being $\Lambda $ the cutoff of our model)\ as indicated
by Eq. (\ref{VEVsinglets}), we find that the SM quarks do not mix with the
heavy exotic quarks and that the mass matrices for up- and down-type SM
quarks are
\begin{eqnarray}
M_{U} &=&\left( 
\begin{array}{ccc}
a_{11}^{\left( U\right) }\lambda ^{8} & 0 & a_{13}^{\left( U\right) }\lambda
^{3} \\ 
0 & a_{22}^{\left( U\right) }\lambda ^{4} & a_{23}^{\left( U\right) }\lambda
^{2} \\ 
0 & 0 & a_{33}^{\left( U\right) }%
\end{array}%
\right) \frac{v}{\sqrt{2}},  \label{Mq} \\
M_{D} &=&\left( 
\begin{array}{ccc}
a_{11}^{\left( D\right) }\lambda ^{7} & a_{12}^{\left( D\right) }\lambda ^{6}
& 0 \\ 
a_{21}^{\left( D\right) }\lambda ^{6} & a_{22}^{\left( D\right) }\lambda ^{5}
& 0 \\ 
0 & 0 & a_{33}^{\left( D\right) }\lambda ^{3}%
\end{array}%
\right) \frac{v}{\sqrt{2}},  \notag
\end{eqnarray}%
where $\lambda =0.225$ is one of the Wolfenstein parameters, $v=246$ GeV the
symmetry breaking scale, and $a_{ij}^{\left( U,D\right) }$ ($i,j=1,2,3$) are $%
\mathcal{O}(1)$ parameters. From the SM quark mass matrix textures given by
Eq. (\ref{Mq}), it follows that the Cabbibo mixing arises from the down-type
quark sector whereas the up quark sector generates the remaining quark
mixing angles. The $\mathcal{O}(1)$ dimensionless couplings $a_{ij}^{\left(
U,D\right) }$ ($i,j=1,2,3$)\ in Eq. (\ref{Mq}) are given by the following
relations:

\begin{eqnarray}
a_{11}^{\left( U\right) } &=&y_{11}^{\left( U\right) }\frac{v_{\rho }}{v},%
\hspace{0.5cm}a_{22}^{\left( U\right) }=y_{11}^{\left( U\right) }\frac{%
v_{\rho }}{v},\hspace{0.5cm}a_{13}^{\left( U\right) }=y_{13}^{\left(
U\right) }\frac{v_{\rho }}{v},  \notag \\
a_{23}^{\left( U\right) } &=&y_{23}^{\left( U\right) }\frac{v_{\rho }}{v},%
\hspace{0.5cm}a_{33}^{\left( U\right) }=y_{33}^{\left( U\right) }\frac{%
v_{\zeta }v_{\eta }}{v\Lambda },  \notag \\
a_{11}^{\left( D\right) } &=&y_{11}^{\left( D\right) }\frac{v_{\eta }}{v},%
\hspace{0.5cm}a_{12}^{\left( D\right) }=y_{12}^{\left( D\right) }\frac{%
v_{\eta }}{v},\hspace{0.5cm}a_{21}^{\left( D\right) }=y_{21}^{\left(
D\right) }\frac{v_{\eta }}{v},  \notag \\
a_{22}^{\left( D\right) } &=&y_{22}^{\left( D\right) }\frac{v_{\eta }}{v},%
\hspace{0.5cm}a_{33}^{\left( D\right) }=y_{33}^{\left( D\right) }\frac{%
v_{\rho }}{v}.  \label{qcouplings}
\end{eqnarray}

Furthermore, we find that the exotic quark masses are

\begin{eqnarray}
m_{T} &=&\lambda y^{\left( T\right) }\frac{v_{\chi }}{\sqrt{2}},  \label{mexotics} \\
m_{J^{1,2}} &=&\lambda y_{1,2}^{\left( J\right) }\frac{v_{\chi }}{\sqrt{2}}=%
\frac{y_{1,2}^{\left( J\right) }}{y^{\left( T\right) }}m_{T}.  \notag
\end{eqnarray}

From Eq. (\ref{Mq}) we find that the up- and down-type SM quark masses are
approximatelly given by

\begin{eqnarray}
m_{u} &\simeq &a_{11}^{\left( U\right) }\lambda ^{8}\frac{v}{\sqrt{2}},%
\hspace{0.5cm}m_{c}\simeq a_{22}^{\left( U\right) }\lambda ^{4}\frac{v}{%
\sqrt{2}},  \notag \\
m_{t} &\simeq &a_{33}^{\left( U\right) }\frac{v}{\sqrt{2}},  \label{mqSM} \\
m_{d} &\simeq &\left\vert a_{11}^{\left( D\right) }a_{22}^{\left( D\right)
}-a_{12}^{\left( D\right) }a_{21}^{\left( D\right) }\right\vert \lambda ^{7}%
\frac{v}{\sqrt{2}},\hspace{1cm}  \notag \\
m_{s} &\simeq &a_{22}^{\left( D\right) }\lambda ^{5}\frac{v}{\sqrt{2}},%
\hspace{0.5cm}m_{b}\simeq a_{33}^{\left( D\right) }\lambda ^{3}\frac{v}{%
\sqrt{2}}.  \notag
\end{eqnarray}

We also find that the CKM quark mixing matrix is approximatelly given by

\begin{eqnarray}
&&V_{CKM}  \label{CKM} \\
&\simeq &\left( 
\begin{array}{ccc}
c_{12}c_{13} & c_{13}s_{12} & e^{i\delta }s_{13} \\ 
e^{-i\delta }c_{12}s_{13}s_{23}-c_{23}s_{12} & c_{12}c_{23}+e^{-i\delta
}s_{12}s_{13}s_{23} & -c_{13}s_{23} \\ 
-s_{12}s_{23}-e^{-i\delta }c_{12}c_{23}s_{13} & c_{12}s_{23}-e^{-i\delta
}c_{23}s_{12}s_{13} & c_{13}c_{23}%
\end{array}%
\right) \allowbreak ,  \notag
\end{eqnarray}

where $c_{ij}=\cos \theta _{ij}$, $s_{ij}=\sin \theta _{ij}$ (with $i\neq j$
and$\ i,j=1,2,3$), $\theta _{ij}$ and $\delta $ being the quark mixing
angles and the CP violating phase, respectively. The quark mixing angles and
the CP violating phase are given by

\begin{eqnarray}
\tan \theta _{12} &\simeq &\frac{a_{12}^{\left( D\right) }}{a_{22}^{\left(
D\right) }}\lambda ,\hspace{1cm}\tan \theta _{23}\simeq \frac{a_{23}^{\left(
U\right) }}{a_{33}^{\left( U\right) }}\lambda ^{2},  \label{quarkangles} \\
\tan \theta _{13} &\simeq &\frac{\left\vert a_{13}^{\left( U\right)
}\right\vert }{a_{33}^{\left( U\right) }}\lambda ^{3},\hspace{1cm}\delta
=-\arg \left( a_{13}^{\left( U\right) }\right) .  \notag
\end{eqnarray}

Here we assume that the $\mathcal{O}(1)$ dimensionless couplings $a_{ij}^{\left( U,D\right) }$ ($i,j=1,2,3$)\ in Eq. (\ref{Mq}) are real except for $a_{13}^{\left( U\right) }$. It is noteworthy
that Eqs. (\ref{mqSM})-(\ref{quarkangles}) give an elegant description of
the SM quark masses and mixing angles in terms of the Wolfenstein parameter $%
\lambda =0.225$ and of parameters of order unity. It is worth commenting
that the observables in the quark sector are connected with the electroweak
symmetry breaking scale $v=246$ GeV through their power dependence on the
Wolfenstein parameter $\lambda =0.225$, with $\mathcal{O}(1)$ coefficients.

The Wolfenstein parameterization \cite{Wolfenstein:1983yz} of the CKM matrix
is given by: 
\begin{equation}
V_{W}\simeq \left( 
\begin{array}{ccc}
1-\frac{\lambda ^{2}}{2} & \lambda & A\lambda ^{3}(\rho -i\eta ) \\ 
-\lambda & 1-\frac{\lambda ^{2}}{2} & A\lambda ^{2} \\ 
A\lambda ^{3}(1-\rho -i\eta ) & -A\lambda ^{2} & 1%
\end{array}%
\right) ,  \label{wolf}
\end{equation}%
with 
\begin{eqnarray}
\lambda &=&0.22535\pm 0.00065,\quad \quad A=0.811_{-0.012}^{+0.022}, \\
\quad \overline{{\rho }} &=&0.131_{-0.013}^{+0.026},\quad \quad \overline{{%
\eta }}=0.345_{-0.014}^{+0.013}, \\
\overline{{\rho }} &\simeq &\rho \left( 1-\frac{{\lambda }^{2}}{2}\right)
,\quad \quad \overline{{\eta }}\simeq \eta \left( 1-\frac{{\lambda }^{2}}{2}%
\right) .
\end{eqnarray}%
The comparison with Eq. (\ref{wolf}) leads to the following relations: 
\begin{eqnarray}
a_{33}^{\left( U\right) } &\simeq &1,\quad \quad a_{23}^{\left( U\right)
}\simeq 0.81,\quad \quad a_{13}^{\left( U\right) }\simeq -0.3e^{i\delta
},\quad \quad \delta =67^{\circ },  \notag \\
a_{22}^{\left( U\right) } &\simeq &\frac{m_{c}}{\lambda ^{4}m_{t}}\simeq
1.43,\quad \quad a_{11}^{\left( U\right) }\simeq \frac{m_{u}}{\lambda
^{8}m_{t}}\simeq 1.27,
\end{eqnarray}%
then it follows that $a_{13}^{\left( U\right) }$ is required to be complex,
as previously assumed and its magnitude is a bit smaller than the remaining $%
\mathcal{O}(1)$ coefficients.

Assuming that the hierarchy of the SM quark masses and quark mixing matrix
elements arises from the $Z_{4}$ and $Z_{12}$ symmetries, we set $a_{21}^{\left( D\right) }=a_{22}^{\left( D\right) }$. We fit the parameters $a_{ij}^{\left( D\right) }$\ ($i\neq j$)\ in Eq. (\ref{Mq}) to reproduce the down-type quark masses and quark mixing parameters. The results are shown in
Table \ref{Observables} for the following best-fit values: 
\begin{equation}
a_{11}^{\left( D\right) }\simeq 0.84,\quad a_{12}^{\left( D\right) }\simeq
0.4,\quad a_{22}^{\left( D\right) }\simeq 0.57,\quad a_{33}^{\left( D\right)
}\simeq 1.42.
\end{equation}

The obtained quark masses and CKM parameters are consistent with the
experimental data. The values of these observables as well as the quark
masses together with the experimental data are shown in Table \ref{Observables}. The experimental values of the quark masses, which are given
at the $M_{Z}$ scale, have been taken from Ref. \cite{Bora:2012tx} (which
are similar to those in \cite{Xing:2007fb}), whereas the experimental values
of the CKM matrix elements and the Jarlskog invariant $J$ are taken from
Ref. \cite{PDG}. As seen from Table \ref{Observables}, all observables in the quark sector are in excellent agreement with the experimental data, excepting $\bigl|V_{td}\bigr|$, which turns out to be larger by a factor $\sim 1.3$ than its corresponding experimental value, and naively deviated 8 sigma away from it.
\begin{table}[tbh]
\begin{center}
\begin{tabular}{c|l|l}
\hline\hline
Observable & Model value & Experimental value \\ \hline
$m_{u}(MeV)$ & \quad $1.47$ & \quad $1.45_{-0.45}^{+0.56}$ \\ \hline
$m_{c}(MeV)$ & \quad $641$ & \quad $635\pm 86$ \\ \hline
$m_{t}(GeV)$ & \quad $172.2$ & \quad $172.1\pm 0.6\pm 0.9$ \\ \hline
$m_{d}(MeV)$ & \quad $2.2$ & \quad $2.9_{-0.4}^{+0.5}$ \\ \hline
$m_{s}(MeV)$ & \quad $60.0$ & \quad $57.7_{-15.7}^{+16.8}$ \\ \hline
$m_{b}(GeV)$ & \quad $2.82$ & \quad $2.82_{-0.04}^{+0.09}$ \\ \hline
$\bigl|V_{ud}\bigr|$ & \quad $0.97419$ & \quad $0.97427\pm 0.00015$ \\ \hline
$\bigl|V_{us}\bigr|$ & \quad $0.22572$ & \quad $0.22534\pm 0.00065$ \\ \hline
$\bigl|V_{ub}\bigr|$ & \quad $0.00351$ & \quad $%
0.00351_{-0.00014}^{+0.00015} $ \\ \hline
$\bigl|V_{cd}\bigr|$ & \quad $0.22548$ & \quad $0.22520\pm 0.00065$ \\ \hline
$\bigl|V_{cs}\bigr|$ & \quad $0.97338$ & \quad $0.97344\pm 0.00016$ \\ \hline
$\bigl|V_{cb}\bigr|$ & \quad $0.0411$ & \quad $0.0412_{-0.0005}^{+0.0011}$
\\ \hline
$\bigl|V_{td}\bigr|$ & \quad $0.0110$ & \quad $0.00867_{-0.00031}^{+0.00029} 
$ \\ \hline
$\bigl|V_{ts}\bigr|$ & \quad $0.0398$ & \quad $0.0404_{-0.0005}^{+0.0011}$
\\ \hline
$\bigl|V_{tb}\bigr|$ & \quad $0.999147$ & \quad $%
0.999146_{-0.000046}^{+0.000021}$ \\ \hline
$J$ & \quad $2.96\times 10^{-5}$ & \quad $(2.96_{-0.16}^{+0.20})\times
10^{-5}$ \\ \hline
$\delta $ & \quad $68^{\circ }$ & \quad $68^{\circ }$ \\ \hline\hline
\end{tabular}%
\end{center}
\caption{Model and experimental values of the quark masses and CKM
parameters.}
\label{Observables}
\end{table}

\section{The scalar potential}

\label{scalarpotential} To build a $SU\left( 3\right) _{C}\otimes SU\left(
3\right) _{L}\otimes U\left( 1\right) _{X}\otimes S_{3}$ invariant scalar
potential it is necessary to decompose the direct product of $S_{3}$
representations into irreducible $S_{3}$ representations. The $S_{3}$ group
has three irreducible representations that can be characterized by their
dimension, i.e., $\mathbf{2}$, $\mathbf{1}$ and $\mathbf{1^{\prime }}$. With
the multiplication rules of the $S_{3}$ group given in the appendix, we have
to assign to the scalar fields in the $S_{3}$ irreps and build the
corresponding scalar potential invariant under the symmetry group.

Since all singlet scalars acquire VEVs at a scale much larger than $v_{\chi
} $, they are very heavy and thus the mixing between these scalar singlets
and the $SU\left( 3\right) _{L}$ scalar triplets can be neglected. For the
sake of simplicity we assume a CP scalar potential with only real couplings
as done in Refs. \cite{M-O,Machado:2010uc}. Then, the renormalizable low
energy scalar potential of the model is constructed with the $S_{3}$ doublet 
$\Phi =\left( \eta ,\chi \right) $ and the non-trivial $S_{3}$ singlet $\rho 
$ fields, in the way invariant under the group $SU(3)_{C}\otimes SU\left(
3\right) _{L}\otimes U\left( 1\right) _{X}\otimes S_{3}$. The renormalizable
low energy scalar potential is given by:%
\begin{align}
V_{H}& =\mu _{\rho }^{2}(\rho ^{\dagger }\rho )+\mu _{\Phi }^{2}\left( \Phi
^{\dagger }\Phi \right) _{\mathbf{1}}+\lambda _{1}(\rho ^{\dagger }\rho
)(\rho ^{\dagger }\rho )  \notag \\
& +\lambda _{2}\left( \Phi ^{\dagger }\Phi \right) _{\mathbf{1}}\left( \Phi
^{\dagger }\Phi \right) _{\mathbf{1}}+\lambda _{3}\left( \Phi ^{\dagger
}\Phi \right) _{\mathbf{1}^{\prime }}\left( \Phi ^{\dagger }\Phi \right) _{%
\mathbf{1}^{\prime }}  \notag \\
& +\lambda _{4}\left( \Phi ^{\dagger }\Phi \right) _{\mathbf{2}}\left( \Phi
^{\dagger }\Phi \right) _{\mathbf{2}}+\lambda _{5}(\rho ^{\dagger }\rho
)\left( \Phi ^{\dagger }\Phi \right) _{\mathbf{1}}  \notag \\
& +\lambda _{6}\left( (\rho ^{\dagger }\Phi )\left( \Phi ^{\dagger }\rho
\right) \right) _{\mathbf{1}}+f\left[ \varepsilon ^{ijk}\left( \Phi _{i}\Phi
_{j}\right) _{\mathbf{1}^{\prime }}\rho _{k}+h.c\right] ,  \label{5}
\end{align}%
where $\Phi _{i}=\left( \eta _{i},\chi _{i}\right) $ is a $S_{3}$ doublet
with $i=1,2,3$.

The $S_{3}$ symmetry in the quadratic term of the scalar potential is softly
broken because the vacuum expectation values of the scalar fields $\eta $
and $\chi $ contained in the $S_{3}$ doublet $\Phi $ satisfy the hierarchy $%
v_{\chi }\gg v_{\eta }$. %
Then, we include the quadratic $S_{3}$ soft-breaking terms $\left( \mu
_{\eta }^{2}-\mu _{\chi }^{2}\right) \left( \eta ^{\dagger }\eta \right) $
and $\mu _{\eta \chi }^{2}\left( \chi ^{\dagger }\eta \right) +h.c$ as done
in Ref. \cite{catano}, and use the $S_{3}$ multiplication rules to rewrite
the low energy scalar potential as follows: 
\begin{align}
V_{H}& =\mu _{\rho }^{2}(\rho ^{\dagger }\rho )+\mu _{\eta }^{2}\left( \eta
^{\dagger }\eta \right) +\mu _{\chi }^{2}\left( \chi ^{\dagger }\chi \right)
+\mu _{\eta \chi }^{2}\left[ \left( \chi ^{\dagger }\eta \right) +\left(
\eta ^{\dagger }\chi \right) \right]  \notag \\
& +\lambda _{1}(\rho ^{\dagger }\rho )^{2}+\left( \lambda _{2}+\lambda
_{4}\right) \left[ \left( \chi ^{\dagger }\chi \right) ^{2}+(\eta ^{\dagger
}\eta )^{2}\right]  \notag \\
& +\lambda _{5}\left[ (\rho ^{\dagger }\rho )(\chi ^{\dagger }\chi )+(\rho
^{\dagger }\rho )(\eta ^{\dagger }\eta )\right]  \label{V1} \\
& +2\left( \lambda _{2}-\lambda _{4}\right) \left( \chi ^{\dagger }\chi
\right) \left( \eta ^{\dagger }\eta \right) +2\left( \lambda _{4}-\lambda
_{3}\right) \left( \chi ^{\dagger }\eta \right) \left( \eta ^{\dagger }\chi
\right)  \notag \\
& +\lambda _{6}\left[ \left( \chi ^{\dagger }\rho \right) (\rho ^{\dagger
}\chi )+\left( \eta ^{\dagger }\rho \right) (\rho ^{\dagger }\eta )\right] 
\notag \\
& +\left( \lambda _{3}+\lambda _{4}\right) \left[ \left( \chi ^{\dagger
}\eta \right) ^{2}+\left( \eta ^{\dagger }\chi \right) ^{2}\right] +2f\left(
\varepsilon ^{ijk}\eta _{i}\chi _{j}\rho _{k}+h.c\right) .  \notag
\end{align}%
It is noteworthy that the $S_{3}$ soft-breaking term $\mu _{\eta \chi
}^{2}\left( \chi ^{\dagger }\eta \right) +h.c$ does not play an important
role neither for the minimization of the scalar potential nor for the
generation of the physical scalar masses Ref. \cite{catano}.

From the previous expressions and from the scalar potential minimization
conditions, the following relations are obtained: 

\begin{eqnarray}
-\mu _{\chi }^{2}&=&\left( \lambda _{2}+\lambda _{4}\right)
v _{\chi }^{2}+\frac{\lambda _{5}}{2}v_{\rho
}^{2}+\left( \lambda _{2}-\lambda _{4}\right)v_{\eta }^{2}-%
\sqrt{2}\frac{fv_{\rho }v_{\eta }}{\upsilon _{\chi }} ,  \notag \\
-\mu _{\eta }^{2}&=&\left( \lambda _{2}+\lambda _{4}\right) v_{\eta
}^{2}+\frac{\lambda _{5}}{2}v_{\rho }^{2}+\left( \lambda
_{2}-\lambda _{4}\right)v_{\chi }^{2}-\sqrt{2}\frac{fv_{\chi }v_{\rho }}{v_{\eta}},  \notag \\
-\mu _{\rho }^{2}&=&\lambda _{1}v_{\rho }^{2}+\frac{\lambda _{5}}{2}\left(v _{\chi }^{2}+v_{\eta }^{2}\right) -\sqrt{2}\frac{fv_{\chi }v_{\eta }}{v_{\rho }}.  \label{A01}
\end{eqnarray}%

Considering the quartic scalar couplings of the same order of magnitude, we
find from the previous relations that the trilinear scalar coupling $f$ has
to be of the order of $v_{\chi }$. Furthermore, from Eq. \ref{A01}, we get
the following relation:%
\begin{equation}
\mu _{\chi }^{2}-\mu _{\eta }^{2}+\left( 2\lambda _{4}+\sqrt{2}\frac{v_{\rho
}}{v_{\eta }}\right) \left( \upsilon _{\chi }^{2}-v_{\eta }^{2}\right) =0.
\end{equation}%
The previous relations imply that the negative quadratic couplings should
satisfy $\mu _{\chi }^{2}\sim -v_{\chi }^{2}$ and $\mu _{\rho }^{2}\sim \mu
_{\eta }^{2}\sim -v_{\rho }^{2}\sim -v_{\eta }^{2}\sim -v^{2}$, being $v=246$
GeV. Therefore, the negative quadratic coupling for the $SU(3)_{L}$ scalar
triplet $\chi $ is of the order of its squared VEV. The remaining negative
quadratic couplings are of the order of the squared VEVs of the $SU(3)_{L}$
scalar triplets $\eta $ and $\rho $. 

From the low energy scalar potential given by Eq. (\ref{V1}), we find that
the physical scalar fields at low energies have the following masses: 
\begin{align}
m_{h^{0}}^{2}& \simeq \frac{\left[ 8\lambda _{4}\lambda _{5}v_{\eta
}^{2}v_{\rho }^{2}+16\lambda _{2}\lambda _{4}v_{\eta }^{4}+\left( 4\lambda
_{1}\left( \lambda _{2}+\lambda _{4}\right) -\lambda _{5}^{2}\right) v_{\rho
}^{4}\right] }{4\left( \lambda _{2}+\lambda _{4}\right) \left( v_{\eta
}^{2}+v_{\rho }^{2}\right) },  \notag \\
m_{H_{1}^{0}}^{2}& \simeq \frac{fv_{\chi }}{\sqrt{2}}\left( \frac{v_{\rho }}{%
v_{\eta }}+\frac{v_{\eta }}{v_{\rho }}\right) ,\hspace{1cm}\hspace{1cm}%
\hspace{1cm}  \notag \\
m_{A^{0}}^{2}& \simeq \frac{fv_{\chi }}{\sqrt{2}}\left( \frac{v_{\eta }}{%
v_{\rho }}+\frac{v_{\rho }}{v_{\eta }}\right) ,  \notag \\
m_{H_{2}^{0}}^{2}& =m_{\overline{H}_{2}^{0}}^{2}\simeq 2\lambda _{4}v_{\chi
}^{2}+\sqrt{2}fv_{\chi }\frac{v_{\rho }}{v_{\eta }},\hspace{1cm}  \notag \\
m_{H_{3}^{0}}^{2}& \simeq \left( \lambda _{2}+\lambda _{4}\right) v_{\chi
}^{2},  \notag \\
m_{H_{1}^{\pm }}^{2}& \simeq \sqrt{2}\left( \frac{v_{\rho }}{v_{\eta }}+%
\frac{v_{\eta }}{v_{\rho }}\right) fv_{\chi },\hspace{1cm}\hspace{1cm} 
\notag \\
m_{H_{2}^{\pm }}^{2}& \simeq \frac{\lambda _{6}}{2}\upsilon _{\chi }^{2}+%
\sqrt{2}fv_{\chi }\frac{v_{\eta }}{v_{\rho }},\hspace{1cm}  \notag \\
m_{G_{1}^{0}}^{2}& =m_{G_{2}^{0}}^{2}=m_{\overline{G}%
_{2}^{0}}^{2}=m_{G_{3}^{0}}^{2}=m_{G_{1}^{\pm }}^{2}=m_{G_{2}^{\pm }}^{2}=0.
\end{align}

It is noteworthy that the physical scalar spectrum at low energies of our
model includes: four massive charged Higgs ($H_{1}^{\pm }$, $H_{2}^{\pm }$),
one CP-odd Higgs ($A^{0}$), three neutral CP-even Higgs ($%
h^{0},H_{1}^{0},H_{3}^{0}$) and two neutral Higgs ($H_{2}^{0},\overline{H}%
_{2}^{0}$) bosons. Here we identify the scalar $h^{0}$ with the SM-like $126$
GeV Higgs boson observed at the LHC. Let us note that the neutral Goldstone
bosons $G_{1}^{0}$, $G_{3}^{0}$, $G_{2}^{0}$ , $\overline{G}_{2}^{0}$ are
associated to the longitudinal components of the $Z$, $Z^{\prime }$, $K^{0}$
and $\overline{K}^{0}$gauge bosons, respectively. Besides that, the charged
Goldstone bosons $G_{1}^{\pm }$ and $G_{2}^{\pm }$ are associated to the
longitudinal components of the $W^{\pm }$ and $K^{\pm }$ gauge bosons,
respectively \cite{331-pisano,M-O}.

In Appendix \ref{ap2} we employ the method of Ref. \cite{Maniatis:2006fs} to
show that the low energy scalar potential is stable when the following
conditions are fulfilled:

\begin{eqnarray}
\lambda _{1}>0, &&\hspace{0.3cm}\lambda _{2}>0,\hspace{0.3cm}\lambda _{4}>0,%
\hspace{0.3cm}\lambda _{6}>0,\hspace{0.3cm}f>0,  \notag \\
\lambda _{2}>\lambda _{3}, &&\hspace{0.3cm}\lambda _{2}+\lambda _{4}>0,%
\hspace{0.3cm}\lambda _{5}+\lambda _{6}>0,  \notag \\
\lambda _{1}(\lambda _{2}+\lambda _{4})>\lambda _{5}^{2}, &&\hspace{0.3cm}%
\lambda _{5}+\lambda _{6}>2\sqrt{\lambda _{1}\left( \lambda _{2}+\lambda
_{4}\right) }.
\end{eqnarray}

\section{Conclusions}

\label{conclusions} 
In this paper we proposed a model based on the symmetry group $SU(3)_{C}\otimes SU\left( 3\right)
_{L}\otimes U\left( 1\right) _{X}\otimes S_{3}\otimes Z_{2}\otimes
Z_{4}\otimes Z_{12}$, which is an extension of the 331 model with $\beta =-%
\frac{1}{\sqrt{3}}$ of Ref.\cite{catano}. Our model successfully accounts
for the observed SM quark mass and mixing pattern. The $S_{3}$ and $Z_{2}$
symmetries are crucial for reducing the number of parameters in the Yukawa
terms for the quark sector and decoupling the bottom quark from the light down and strange quarks. The
observed hierarchy of the SM quark masses and quark mixing matrix elements
arises from the $Z_{4}$ and $Z_{12}$ symmetries, which are broken at a very
high scale by the $SU(3)_{L}$ scalar singlets ($\sigma$,$\zeta$) and $\tau $, charged
under these symmetries, respectively. The Cabbibo mixing arises from the
down-type quark sector whereas the up quark sector generates the remaining mixing angles. The SM quark masses are generated from Yukawa terms
involving the $SU(3)_{L}$ scalar triplets $\eta $ and $\rho $, which acquire
VEVs at the electroweak scale $v=246$ GeV. On the other hand, the exotic
quark masses arise from Yukawa terms involving the $SU(3)_{L}$ scalar
triplet $\chi $, which acquires a VEV at the TeV scale. The obtained values
of the quark masses, the magnitudes of the CKM matrix elements, the CP
violating phase, and the Jarlskog invariant are consistent with the
experimental data. The complex phase responsible for CP violation in the
quark sector has been assumed to come from a seven dimensional up-type quark Yukawa term. 

\subsection*{Acknowledgements}

A.E.C.H was supported by Fondecyt (Chile), Grant No. 11130115 and by DGIP
internal Grant No. 111458. R.M. was supported by COLCIENCIAS and by Fondecyt
(Chile), Grant No. 11130115.

\section*{Appendices}

\appendix


\section{: The product rules for $S_{3}$}

\label{ap0} The $S_{3}$ group has three irreducible representations that can
be characterized by their dimension, i.e., $\mathbf{2}$, $\mathbf{1}$ and $%
\mathbf{1^{\prime }}$. Considering two $S_{3}$ doublet representations $%
(x_{1},x_{2})$ and $(y_{1},y_{2})$, the direct product can be decomposed as
follows \cite{Ishimori:2010au}: 
\begin{eqnarray}
\left( 
\begin{array}{c}
x_{1} \\ 
x_{2}%
\end{array}%
\right) _{\mathbf{2}}\otimes \left( 
\begin{array}{c}
y_{1} \\ 
y_{2}%
\end{array}%
\right) _{\mathbf{2}} &=&\left( x_{1}y_{1}+x_{2}y_{2}\right) _{\mathbf{1}%
}+\left( x_{1}y_{2}-x_{2}y_{1}\right) _{\mathbf{1}^{\prime }}  \notag \\
&&+\left( 
\begin{array}{c}
x_{1}y_{1}-x_{2}y_{2} \\ 
x_{1}y_{2}+x_{2}y_{1}%
\end{array}%
\right) _{\mathbf{2}},
\end{eqnarray}%
\begin{eqnarray}
\left( 
\begin{array}{c}
x_{1} \\ 
x_{2}%
\end{array}%
\right) _{\mathbf{2}}\otimes \left( y\right) _{\mathbf{1}^{\prime }}
&=&\left( 
\begin{array}{c}
-x_{2}y \\ 
x_{1}y%
\end{array}%
\right) _{\mathbf{2}},  \notag \\
\left( x\right) _{\mathbf{1}^{\prime }}\otimes \left( y\right) _{\mathbf{1}%
^{\prime }} &=&\left( xy\right) _{\mathbf{1}}.
\end{eqnarray}%
With these multiplication rules we have to assign to the scalar fields in
the $S_{3}$ irreps and build the corresponding scalar potential invariant
under the symmetry group. 



\section{: Stability conditions of the low energy scalar potential}

\label{ap2}

In this subsection we are going to determine the conditions required to have
a stable scalar potential by following the method described in Ref. \cite{Maniatis:2006fs}. The gauge invariant and renormalizable low energy scalar
potential as a function of the fields $\phi _{1}=\chi $, $\phi _{2}=\eta $
and $\phi _{3}=\rho $ is a linear hermitian combination of the following
terms: 
\begin{equation}
\phi _{i}\phi _{j},\;\;\;\;\;\;\phi _{i}\phi _{j}\phi _{k}\phi _{l}
\end{equation}%
%
%
%
%
%
%
%
%
%
%
%
%
%
%
%
%
%
%
%
%
%
%
%
%
%
%
%
%
%
%
%
where $i,j,k,l=\phi _{1}$, $\phi _{2}$ and $\phi _{3}$. To discuss the
stability of the potential, its minimum, and its gauge invariance one can
make the following arrangement of the scalar fields by using $2\times 2$
hermitian matrices as follows: 
\begin{eqnarray}
\widetilde{K}_{(\phi _{i}\phi _{j})} &=&\left( 
\begin{array}{cc}
\phi _{i}^{\dagger }\phi _{i} & \phi _{i}^{\dagger }\phi _{j} \\ 
\phi _{j}^{\dagger }\phi _{i} & \phi _{j}^{\dagger }\phi _{j}%
\end{array}%
\right) ,  \notag \\
&=&\frac{1}{2}\left( K_{0\left( \phi _{i}\phi _{j}\right) }1_{2\times
2}+K_{a\left( \phi _{i}\phi _{j}\right) }\sigma ^{a}\right)
\end{eqnarray}%
%
%
%
%
%
where $(\phi _{i}\phi _{j})=\rho \eta ,\rho \chi ,\eta \chi $, $\sigma ^{a}$
($a=1,2,3$) are the Pauli matrices and $1_{2\times 2}$ is the identity
matrix. From the previous expressions one can build the following bilinear
terms as functions of the scalar fields: 
\begin{eqnarray}
K_{0\left( \phi _{i}\phi _{j}\right) } &=&\phi _{i}^{\dagger }\phi _{i}+\phi
_{j}^{\dagger }\phi _{j},\hspace{1cm}\hspace{1cm}  \notag \\
K_{a\left( \phi _{i}\phi _{j}\right) } &=&\sum_{i,j}\left( \phi
_{i}^{\dagger }\phi _{j}\right) \sigma _{ij}^{a}.
\end{eqnarray}%
\quad The properties of the potential can be analyzed in terms of $%
K_{0\left( \phi _{i}\phi _{j}\right) }$ and $\vec{K}_{\left( \phi _{i}\phi
_{j}\right) }$ with $\phi _{i}\phi _{j}=\rho \eta ,\rho \chi ,\eta \chi $ in
the domain $K_{0}\geq 0$ y $K_{0}^{2}\geq \vec{K}^{2}$. Defining $\vec{\kappa%
}=\vec{K}/K_{0}$ the potential can be written as
\begin{eqnarray}
V &=&V_{2}+V_{4},  \notag \\
V_{2} &=&\sum_{(\phi _{i}\phi _{j})}K_{0(\phi _{i}\phi _{j})}\vec{J}_{2(\phi
_{i}\phi _{j})}(\vec{\kappa}),\;\;\;\;  \notag \\
\vec{J}_{2{(\phi _{i}\phi _{j})}}(\vec{\kappa}) &=&\xi _{0(\phi _{i}\phi
_{j})}+\vec{\xi}_{(\phi _{i}\phi _{j})}^{T}\vec{\kappa}_{(\phi _{i}\phi
_{j})},  \notag \\
V_{4} &=&\sum_{(\phi _{i}\phi _{j})}K_{0(\phi _{i}\phi _{j})}^{2}\vec{J}%
_{4(\phi _{i}\phi _{j})}(\vec{\kappa}),\;\;\;  \label{potential} \\
\;\vec{J}_{4{(\phi _{i}\phi _{j})}}(\vec{\kappa}) &=&\eta _{00(\phi _{i}\phi
_{j})}+2\vec{\eta}_{(\phi _{i}\phi _{j})}^{T}\vec{\kappa}_{(\phi _{i}\phi
_{j})}  \notag \\
&&+\vec{\kappa}_{(\phi _{i}\phi _{j})}^{T}E_{(\phi _{i}\phi _{j})}\vec{\kappa%
}_{(\phi _{i}\phi _{j})},  \notag
\end{eqnarray}%
where $E_{(\phi _{i}\phi _{j})}$ is a $3\times 3$ matrix and the functions $J_{2{(\phi _{i}\phi _{j})}}(\vec{\kappa})$ and $J_{4{(\phi _{i}\phi _{j})}}(%
\vec{\kappa})$ are defined in the domain $|\vec{\kappa}|\leq 1$. The
stability of the scalar potential requires that it has to be bounded from
below. The stability is determined from the behavior of $V$ in the limit $%
K_{0}\rightarrow \infty $, i.e., 
\begin{equation}
J_{4{(\phi _{i}\phi _{j})}}(\vec{\kappa})\geq 0,
\end{equation}%
for all $|\vec{\kappa}|\leq 1$. To impose $J_{4{(\phi _{i}\phi _{j})}}(\vec{%
\kappa})$ to be positively defined it is enough to consider the values of
all stationary points in the domain $|\kappa |<1$ and $|\kappa |=1$. This
results in a bound for $\eta _{00{(\phi _{i}\phi _{j})}}$, $\vec{\eta}_{0{%
(\phi _{i}\phi _{j})}}$ and $E{(\phi _{i}\phi _{j})}$, which parametrize the
quartic terms of the potential included in $V_{4}$.

For $|\vec{\kappa}|<1$ the stationary points should satisfy 
\begin{equation}
E\vec{\kappa}_{(\phi _{i}\phi _{j})}=-\vec{\eta}_{(\phi _{i}\phi
_{j})},\;\;\;\;\;\;|\vec{\kappa}|<1.
\end{equation}

For the case where $\det E\neq 0$, the following relation is obtained: 
\begin{equation}
\;J_{4(\phi _{i}\phi _{j})}(\vec{\kappa})|_{est}=\eta _{00{(\phi _{i}\phi
_{j})}}-\vec{\eta}_{(\phi _{i}\phi _{j})}^{T}E^{-1}\vec{\eta}_{(\phi
_{i}\phi _{j})}.
\end{equation}

For $|\vec{\kappa}|=1$ the stationary points are obtained from the function: 
\begin{equation}
F_{4(\phi _{i}\phi _{j})}(\vec{\kappa})=J_{4(\phi _{i}\phi _{j})}(\kappa
)+u(1-\vec{\kappa}^{2}),
\end{equation}%
where $u$ is a Lagrange multiplier that satisfies the following condition 
\begin{eqnarray}
(E_{(\phi _{i}\phi _{j})}-u)\vec{\kappa} &=&-\vec{\eta}_{(\phi _{i}\phi
_{j})},  \notag \\
\;J_{4(\phi _{i}\phi _{j})}(\vec{\kappa})|_{est} &=&u+\eta _{00{(\phi
_{i}\phi _{j})}} \\
&&-\vec{\eta}_{(\phi _{i}\phi _{j})}^{T}(E_{(\phi _{i}\phi _{j})}-u)^{-1}%
\vec{\eta}_{(\phi _{i}\phi _{j})}.  \notag
\end{eqnarray}%
The stationary points of $J_{4(\phi _{i}\phi _{j})}(\kappa )$ for $|\kappa
|\leq 1$ can be obtained from: 
\begin{eqnarray}
f_{(\phi _{i}\phi _{j})}(u) &=&J_{4{(\phi _{i}\phi _{j})}}(\vec{\kappa}%
)|_{est}>0,  \notag \\
f_{(\phi _{i}\phi _{j})}^{\prime }\left( u\right) &>&0.
\label{stabilityconditions}
\end{eqnarray}%
Considering that the quartic terms of the scalar potential are dominant
when the vacuum expectation values of the scalar fields take large values,
these terms will be the most relevant to analyze the stability of the scalar
potential. Following the method described in Ref. \cite{Maniatis:2006fs}, we
proceed to rewrite the quartic terms of the scalar potential in terms of
bilinear combinations of the scalar fields. To this end, the bilinear
combinations of the scalar fields are included in the following matrices: 
\begin{eqnarray}
\widetilde{K}_{\rho \eta } &=&\left( 
\begin{array}{cc}
\rho ^{\dagger }\rho & \eta ^{\dagger }\rho \\ 
\rho ^{\dagger }\eta & \eta ^{\dagger }\eta%
\end{array}%
\right) =\frac{1}{2}\left( K_{0\left( \rho \eta \right) }1_{2\times
2}+K_{a\left( \rho \eta \right) }\sigma ^{a}\right) ,  \notag \\
\widetilde{K}_{\rho \chi } &=&\left( 
\begin{array}{cc}
\rho ^{\dagger }\rho & \chi ^{\dagger }\rho \\ 
\rho ^{\dagger }\chi & \chi ^{\dagger }\chi%
\end{array}%
\right) =\frac{1}{2}\left( K_{0\left( \rho \chi \right) }1_{2\times
2}+K_{a\left( \rho \chi \right) }\sigma ^{a}\right) ,  \notag \\
\widetilde{K}_{\eta \chi } &=&\left( 
\begin{array}{cc}
\eta ^{\dagger }\eta & \chi ^{\dagger }\eta \\ 
\eta ^{\dagger }\chi & \chi ^{\dagger }\chi%
\end{array}%
\right) =\frac{1}{2}\left( K_{0\left( \eta \chi \right) }1_{2\times
2}+K_{a\left( \eta \chi \right) }\sigma ^{a}\right) ,  \notag \\
&&  \label{S1}
\end{eqnarray}%
where $\sigma ^{a}$ ($a=1,2,3$) are the Pauli matrices and $1_{2\times 2}$
is the $2\times 2$ identity matrix. From the previous expressions, we find
that the bilinear combinations of the scalar fields appearing in Eq. (\ref%
{S1}) are given by: 
\begin{eqnarray}
K_{0\left( \rho \eta \right) } &=&\rho ^{\dagger }\rho +\eta ^{\dagger }\eta
,\hspace{0.5cm}K_{0\left( \rho \chi \right) }=\rho ^{\dagger }\rho +\chi
^{\dagger }\chi ,\hspace{0.5cm}  \notag \\
K_{0\left( \eta \chi \right) } &=&\eta ^{\dagger }\eta +\chi ^{\dagger }\chi
,  \label{S2} \\
K_{a\left( \rho \eta \right) } &=&\left( \rho ^{\dagger }\rho \right) \sigma
_{11}^{a}+\left( \eta ^{\dagger }\eta \right) \sigma _{22}^{a}+\left( \rho
^{\dagger }\eta \right) \sigma _{12}^{a}+\left( \eta ^{\dagger }\rho \right)
\sigma _{21}^{a},  \notag \\
K_{a\left( \rho \chi \right) } &=&\left( \rho ^{\dagger }\rho \right) \sigma
_{11}^{a}+\left( \chi ^{\dagger }\chi \right) \sigma _{22}^{a}+\left( \rho
^{\dagger }\chi \right) \sigma _{12}^{a}+\left( \chi ^{\dagger }\rho \right)
\sigma _{21}^{a},  \notag \\
K_{a\left( \eta \chi \right) } &=&\left( \eta ^{\dagger }\eta \right) \sigma
_{11}^{a}+\left( \chi ^{\dagger }\chi \right) \sigma _{22}^{a}+\left( \eta
^{\dagger }\chi \right) \sigma _{12}^{a}+\left( \chi ^{\dagger }\eta \right)
\sigma _{21}^{a}.  \notag
\end{eqnarray}

Since the stability of the scalar potential is determined from its quartic
terms, the stationary solutions consistent with a stable scalar potential
are described by the following functions: 
\begin{eqnarray}
f_{\rho \eta }\left( u\right) &=&u+E_{00\left( \rho \eta \right)
}-E_{a\left( \rho \eta \right) }\left( E_{\left( \rho \eta \right)
}-u1_{3\times 3}\right) _{ab}^{-1}E_{b\left( \rho \eta \right) },  \notag \\
f_{\rho \chi }\left( u\right) &=&u+E_{00\left( \rho \chi \right)
}-E_{a\left( \rho \chi \right) }\left( E_{\left( \rho \chi \right)
}-u1_{3\times 3}\right) _{ab}^{-1}E_{b\left( \rho \chi \right) },  \notag \\
f_{\eta \chi }\left( u\right) &=&u+E_{00\left( \eta \chi \right)
}-E_{a\left( \eta \chi \right) }\left( E_{\left( \eta \chi \right)
}-u1_{3\times 3}\right) _{ab}^{-1}E_{b\left( \eta \chi \right) },  \notag \\
&&  \label{S5}
\end{eqnarray}%
where, for the $\rho $ and $\eta $ fields, we have 
\begin{eqnarray}
E_{00\left( \rho \eta \right) } &=&\frac{\lambda _{1}+\lambda _{2}+\lambda
_{4}+\lambda _{5}}{4},\hspace{1cm}  \notag \\
E_{a\left( \rho \eta \right) } &=&\frac{\lambda _{1}-\lambda _{2}-\lambda
_{4}}{4}\delta _{a3},  \notag \\
E_{\left( \rho \eta \right) } &=&\frac{1}{4}\left( 
\begin{array}{ccc}
\lambda _{6} & 0 & 0 \\ 
0 & \lambda _{6} & 0 \\ 
0 & 0 & \lambda _{1}+\lambda _{2}+\lambda _{4}-\lambda _{5}%
\end{array}%
\right) ,  \label{MS1}
\end{eqnarray}

In the same manner, for the multiplets $\rho $ and $\chi $, the expressions
are 
\begin{eqnarray}
E_{00\left( \rho \chi \right) } &=&\frac{\lambda _{1}+\lambda _{2}+\lambda
_{4}+\lambda _{5}}{4},\hspace{1cm}  \notag \\
E_{a\left( \rho \chi \right) } &=&\frac{\lambda _{1}-\lambda _{2}-\lambda
_{4}}{4}\delta _{a3},  \notag \\
E_{\left( \rho \chi \right) } &=&\frac{1}{4}\left( 
\begin{array}{ccc}
\lambda _{6} & 0 & 0 \\ 
0 & \lambda _{6} & 0 \\ 
0 & 0 & \lambda _{1}+\lambda _{2}+\lambda _{4}-\lambda _{5}%
\end{array}%
\right) .  \label{MS2}
\end{eqnarray}%
Similarly, for the $\eta $ and $\chi $ fields, we find: 
\begin{equation}
E_{00\left( \eta \chi \right) }=\lambda _{2},\hspace{0.5cm}E_{a\left( \eta
\chi \right) }=0,\hspace{0.5cm}E_{\left( \eta \chi \right) }=\left( 
\begin{array}{ccc}
\lambda _{4} & 0 & 0 \\ 
0 & -\lambda _{3} & 0 \\ 
0 & 0 & \lambda _{4}%
\end{array}%
\right) .  \label{MS3}
\end{equation}

Following Ref. \cite{Maniatis:2006fs}, we determine the stability of the
scalar potential from the conditions: 
\begin{equation}
f_{\rho \eta }\left( u\right) >0,\hspace{0.5cm}f_{\rho \chi }\left( u\right)
>0,\hspace{0.5cm}f_{\eta \chi }\left( u\right) >0.  \label{S6}
\end{equation}

We use the theorem of stability of the scalar potential of Ref. \cite{Maniatis:2006fs} to determine the stability conditions of the scalar
potential. To this end, the condition $f_{\rho \eta }\left( u\right) >0$ is
analyzed for the set of values of $u$ which include the $0$, (since $f%
{\acute{}}%
_{\rho \eta }\left( 0\right) >0$) the roots $u_{\rho \eta }^{\left( 1\right)
}$ and $u_{\rho \eta }^{\left( 2\right) }$ of the equation $f%
{\acute{}}%
_{\rho \eta }\left( u\right) =0$ and the eigenvalues $\widetilde{E}_{\left(
\rho \eta \right) }^{\left( a\right) }$\ of the matrix $E_{\left( \rho \eta
\right) }$ where $f_{\rho \eta }\left( \widetilde{E}_{\left( \rho \eta
\right) }^{\left( a\right) }\right) $ is finite and $f%
{\acute{}}%
_{\rho \eta }\left( \widetilde{E}_{\left( \rho \eta \right) }^{\left(
a\right) }\right) \geq 0$ . We proceed in a similar way when analyzing the
conditions $f_{\rho \chi }\left( u\right) >0$ and $f_{\eta \chi }\left(
u\right) >0$.

Therefore, the scalar potential is stable when the following conditions are
fulfilled: 
\begin{eqnarray}
\lambda _{1} &>&0,\hspace{1cm}\lambda _{2}>0,\hspace{1cm}\lambda _{6}>0,%
\hspace{1cm}\lambda _{2}>\lambda _{3},  \notag \\
\lambda _{2}+\lambda _{4} &>&0,\hspace{1cm}\lambda _{5}+\lambda _{6}>2\sqrt{%
\lambda _{1}\left( \lambda _{2}+\lambda _{4}\right) }.  \label{S7a}
\end{eqnarray}%
%
%

From the minimization conditions of the scalar potential, the stability of
the scalar potential and the Higgs masses we can find other restrictions
for the quartic couplings of the scalar potentials. Having masses $%
m_{H_{1}^{\pm }}^{2}$, $m_{H_{1}^{0}}^{2}$ and $m_{H_{1}^{0}}^{2}$
positively defined requires the following condition: 
\begin{equation}
f>0.
\end{equation}%
%
%
In the same manner, the conditions $\lambda _{6}>0$ y $\lambda _{2}+\lambda
_{6}>0$ guarantee that $m_{H_{3}^{0}}^{2}$ and $m_{H_{2}^{\pm }}^{2}$ are positively defined, respectively. From the expressions corresponding to the
masses of the fields $h^{0}$, $H_{2}^{0}$ y $\bar{H}_{2}^{0}$, it is
necessary to impose additional conditions that guarantee that they are
positively defined, i.e., 
\begin{equation}
\lambda _{4}>0,\;\;\;\;\;\;\;\lambda _{1}(\lambda _{2}+\lambda _{4})>\lambda
_{5}^{2}
\end{equation}%
%
%
Then, we get: 
\begin{equation}
\lambda _{5}+\lambda _{6}>0
\end{equation}%
%
%
Finally the stability conditions of the low energy scalar potential can be summarized in the following form:

\begin{eqnarray}
\lambda _{1}>0, &&\hspace{0.3cm}\lambda _{2}>0,\hspace{0.3cm}\lambda _{4}>0,%
\hspace{0.3cm}\lambda _{6}>0,\hspace{0.3cm}f>0,  \notag \\
\lambda _{2}>\lambda _{3}, &&\hspace{0.3cm}\lambda _{2}+\lambda _{4}>0,%
\hspace{0.3cm}\lambda _{5}+\lambda _{6}>0,  \notag \\
\lambda _{1}(\lambda _{2}+\lambda _{4})>\lambda _{5}^{2}, &&\hspace{0.3cm}%
\lambda _{5}+\lambda _{6}>2\sqrt{\lambda _{1}\left( \lambda _{2}+\lambda
_{4}\right) }.
\end{eqnarray}


\end{document}